\documentclass[conference]{IEEEtran}
\IEEEoverridecommandlockouts
\usepackage{cite}
\usepackage{amsmath,amssymb,amsfonts}
\usepackage{algorithmic}
\usepackage{graphicx}
\usepackage{float}
\usepackage{subfigure}
\usepackage{textcomp}
\usepackage{xcolor}
\usepackage{multirow}
\usepackage{makecell}

\def\BibTeX{{\rm B\kern-.05em{\sc i\kern-.025em b}\kern-.08em
    T\kern-.1667em\lower.7ex\hbox{E}\kern-.125emX}}
\begin{document}

\bstctlcite{IEEEexample:BSTcontrol}
\title{RCW-CIM: A Digital CIM-based LLM Accelerator with Read-Compute/Write\\
}
\author{\IEEEauthorblockN{Yan-Cheng Guo\textsuperscript{1}, Tian-Sheuan Chang\textsuperscript{1}, Senior Member, IEEE, and Jian-Wei Su\textsuperscript{2}}
\IEEEauthorblockA{\textit{\textsuperscript{1}{Institute of Electronics, National Yang Ming Chiao Tung University,}}
Hsinchu, Taiwan
}
\IEEEauthorblockA{\textit{\textsuperscript{2}{Industrial Technology Research Institute, }}
Hsinchu, Taiwan
}
}

\maketitle

\begin{abstract}%
Digital computing-in-memory (DCIM) has emerged as a promising solution for large language model (LLM) acceleration by minimizing data transfers between external DRAM and on-chip accelerators while maintaining high precision for superior accuracy. However, existing CIM architectures often overlook weight update latency, which becomes critical as LLM weights are far larger than a single CIM macro’s capacity. To address this issue, this paper proposes a read-compute/write (RCW) architecture that effectively minimizes weight update latency, along with a nonlinear operator fusion that further mitigates dependency-induced latency. The proposed RCW reduces decoding computing latency by 21.59\% on the Llama2-7B model. In addition, the nonlinear operator fusion mechanism achieves a 69.17\% latency reduction through efficient partial accumulation and group-based approximation. Furthermore, a weight-stationary and output-column-stationary (WS-OCS) dataflow is introduced to reduce both external DRAM access and internal CIM weight updates—by 51.6\% and 87.6\%, respectively—during the prefill phase of 1024 tokens, leading to an overall 49.76\% latency reduction. Fabricated using TSMC 22 nm CMOS technology and operating at 100 MHz, the proposed RCW-CIM achieves 3.28 TOPS and 42.3 TOPS/W, enabling 4.2 ms prefill latency and 26.87 decoded tokens per second for the INT4-weight Llama2 model with dual DDR5-6400 memory.

\end{abstract}

\section{Introduction}

Large language models (LLMs) have gained popularity due to their high performance in natural language processing tasks~\cite{2018_gpt1}.
To accommodate fine-tuning for diverse tasks, model sizes have grown substantially, enabling excellent accuracy across a wide range of applications~\cite{2020_gpt3}.
However, the massive model weights significantly increase data transfer between DRAM and accelerators, and lower-precision representations are insufficient to maintain high accuracy. One promising approach is digital computing-in-memory (DCIM) \cite{2021_digital_CIM}, which combines computation and memory to reduce the weight transfer overhead while minimizing computational loss and preserving high precision.

However, prior DCIM work for LLMs~\cite{2025_CELLA} primarily focuses on reducing data alignment latency and encoding for weights or KV caches, without addressing weight update latency on the CIM—a critical performance bottleneck, especially given that LLM weights far exceed the size of a single CIM macro. Similarly, \cite{2025_LLM_CIM} implements nonlinear function computation within the CIM macro, but does not account for global latency of nonlinear operations and supports only 8-bit precision, which is insufficient for accurate nonlinear computation (Fig.~\ref{0_introduction}).

To overcome these limitations, we propose RCW-CIM, which implements a read-compute/write (RCW) to minimize weight update latency and employs operator fusion for high-precision nonlinear functions to reduce global latency while maintaining high accuracy.
RCW-CIM achieves 3.28 TOPS and 42.3 TOPS/W using TSMC 22 nm technology at 100 MHz, with a prefill latency of 4.2 ms and a decoding throughput of 26.87 tokens per second using dual DDR5-6400 memory.

The paper is organized as follows: Section II presents the RCW-CIM architecture and dataflow, Section III shows experimental results and comparisons, and Section IV concludes the paper.

\begin{figure}[tb]
\centering
\includegraphics[width=1.0\linewidth]{figures/0_introduction.png}
\caption{Prior work \cite{2025_LLM_CIM, 2025_CELLA} typically ignores weight-update latency and does not optimize nonlinear-function execution. In contrast, the proposed RCW-CIM adopts a read-compute/write to minimize weight-update latency and supports nonlinear operator fusion to reduce dependencies in Softmax and RMSNorm operations.}
\label{0_introduction}
\end{figure}

\section{Proposed hardware architecture and data flow} 

\subsection{The overall architecture and CIM macro}
\begin{figure}[tb]
\centering
\includegraphics[width=0.9\linewidth]{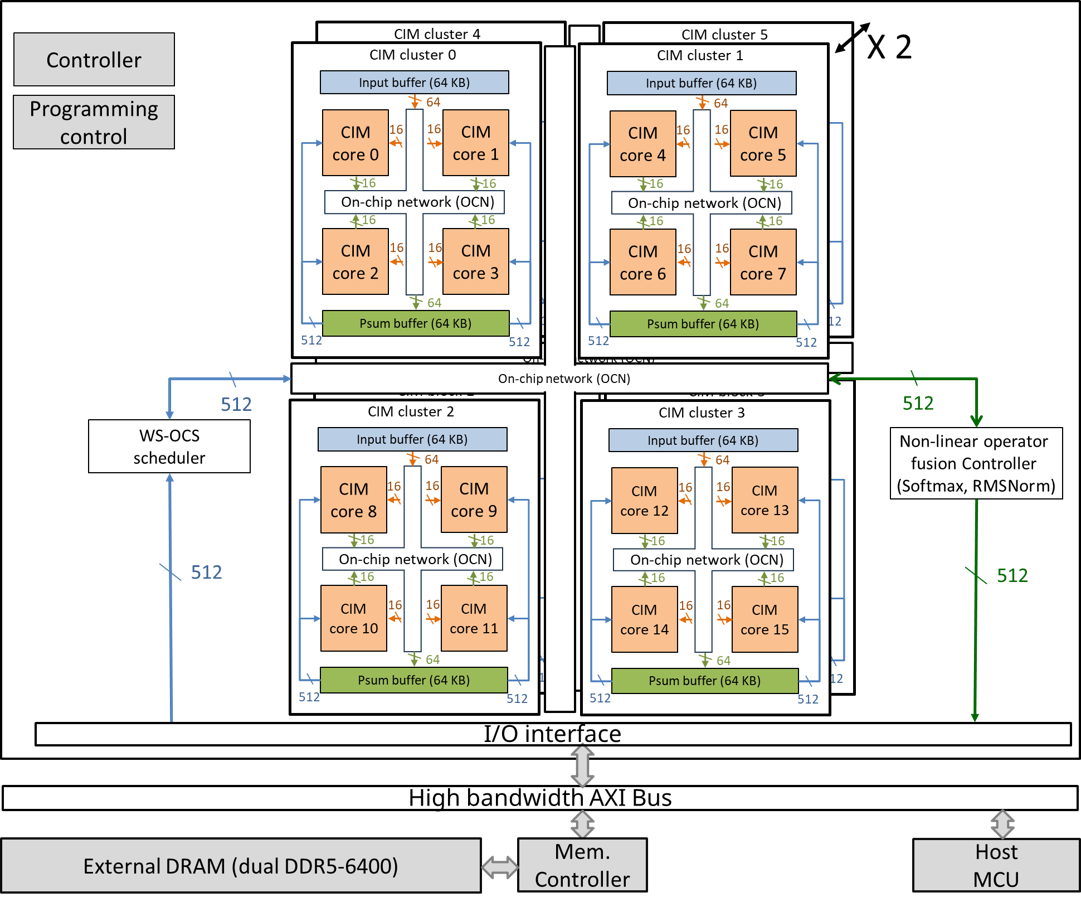}
\caption{The RCW-CIM overall architecture comprises 64 multi-CIM cores, a high-efficiency WS-OCS scheduler, and a nonlinear operator fusion controller supporting Softmax and RMSNorm.}
\label{2_overall_architecture}
\end{figure}

Fig.~\ref{2_overall_architecture} illustrates the system-level architecture of RCW-CIM, which integrates a WS-OCS scheduling unit optimized for weight-update latency, a high-performance nonlinear operator fusion controller, and eight CIM clusters. Each cluster comprises four energy-efficient CIM cores, a 64 KB input-reuse buffer, and a 64 KB partial-sum buffer.

\begin{figure}[tb]
\centering
\includegraphics[width=1.0\linewidth]{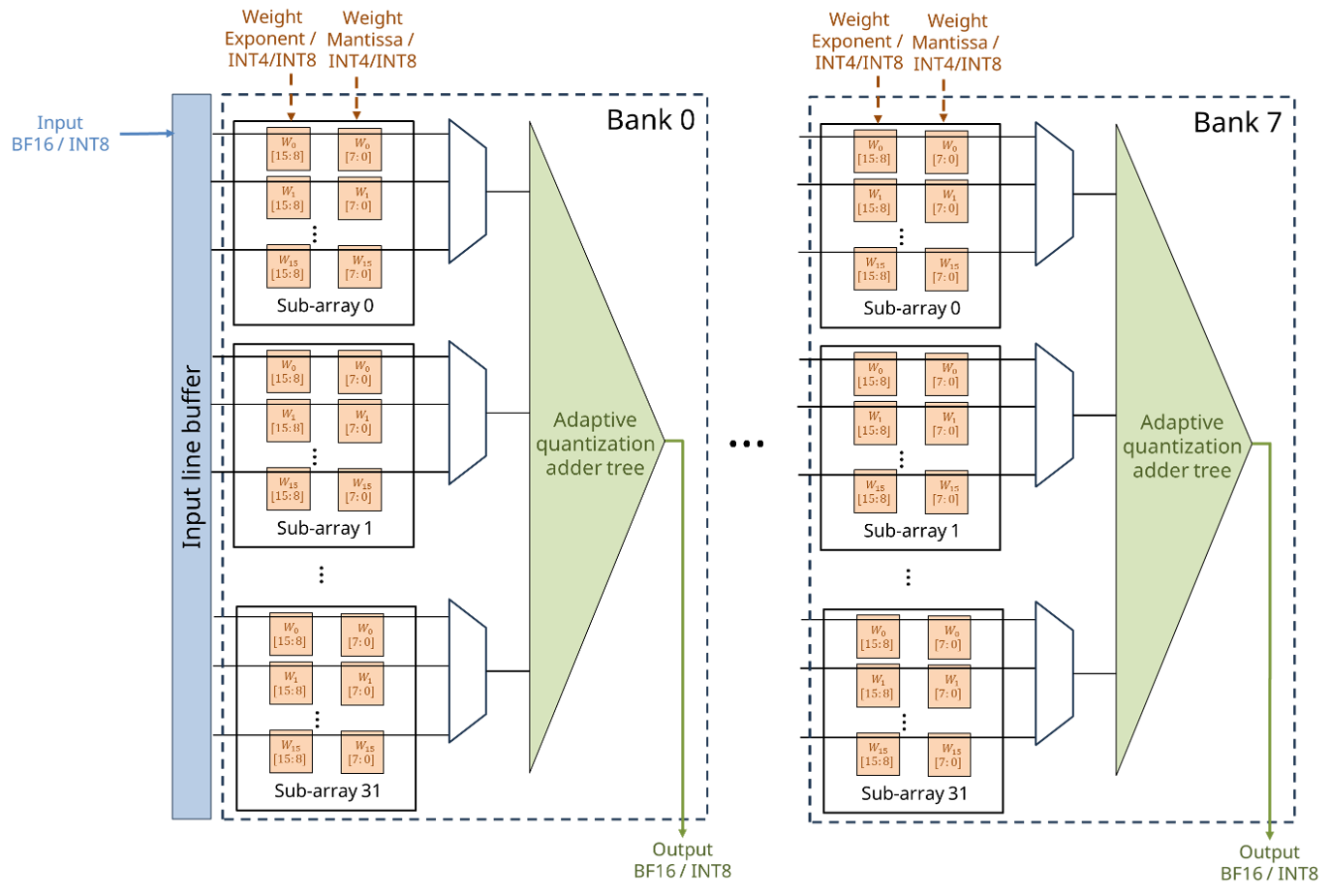}
\caption{The proposed energy-efficient digital SRAM-based CIM macro.}
\label{1_CIM_macro}
\end{figure}

Fig.~\ref{1_CIM_macro} illustrates the proposed energy-efficient digital SRAM-based CIM macro, which supports FP16 precision for nonlinear function computation and a high-throughput dual INT4/INT8 computing mode. 
The CIM macro consists of eight banks, each capable of performing 32 MAC operations in parallel.

\begin{figure}[tb]
\centering
\includegraphics[width=1.0\linewidth]{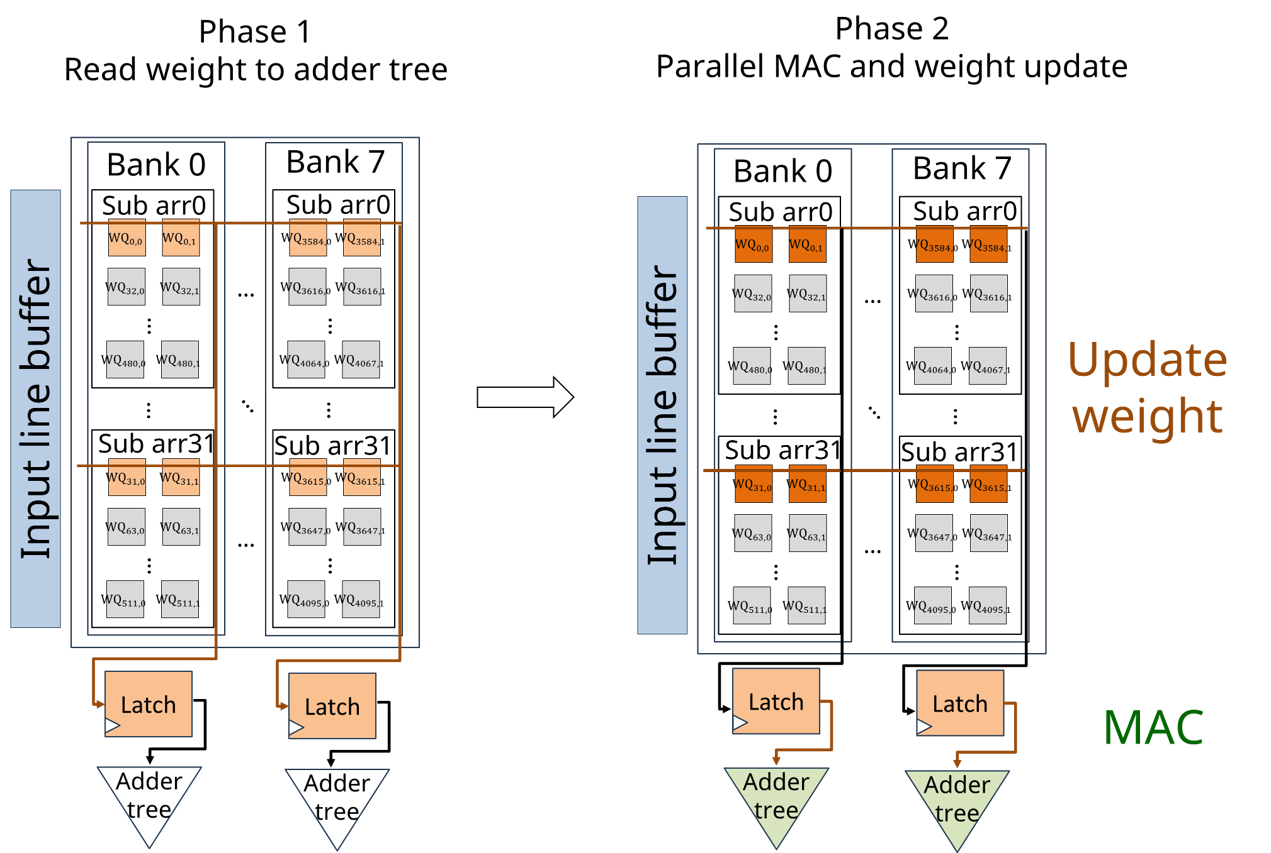}
\caption{Read-compute/write (RCW) in the CIM macro: Phase 1 reads weights into the adder tree, and Phase 2 performs parallel MAC and weight update.}
\label{3_RMW_and_compute_pipeline}
\end{figure}

\subsection{Low weight update latency with read-compute/write}
To minimize LLM weight update latency, we propose a read-compute/write operation for the CIM macro as shown in Fig.~\ref{3_RMW_and_compute_pipeline} to overlap computing and weight update for lower latency.
This performs a two-phase operation: in Phase 1, weights are read and latched for compact adder-tree computation; in Phase 2, MAC operations and weight updates execute concurrently. With this, weight update latency can be hided during computation.
To further reduce WL decoding delay and precharge power, the selected WL remains active throughout both phases, enabling energy-efficient updates. Additionally, this approach supports low-latency transitions between nonlinear function execution and MAC computation via a deeply parallel weight-update and compute design.

\begin{figure}[tb]
\centering
\includegraphics[width=0.8\linewidth]{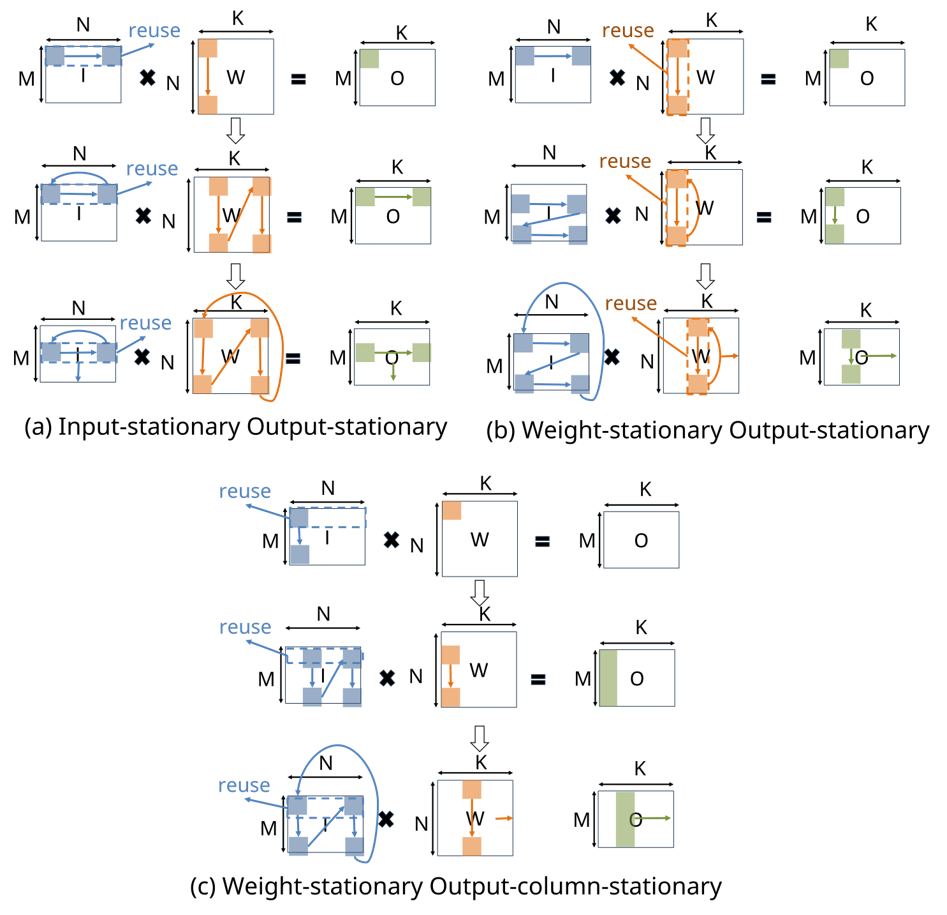}
\caption{Comparative analysis of the IS-OS, WS-OS, and WS-OCS dataflows, where the input, weight, and output matrices have dimensions $M \times N$, $N \times K$, and $M \times K$, respectively. The accelerator processes data in tiles of sizes $m \times n$ for input, $n \times k$ for weight, and $m \times k$ for output. (a) Input-stationary output-stationary dataflow (b) Weight-stationary output-stationary dataflow (c) Proposed Weight-stationary output-column-stationary dataflow}
\label{5_OCS_dataflow}
\end{figure}

\begin{table}[tb]
\caption{DRAM access and CIM update for IS, WS, IS-OS, WS-OS, and WS-OCS dataflows}
\label{tb01_WS_OCS_dataflow}
\centering
\renewcommand{\arraystretch}{1.5} %
\setlength{\tabcolsep}{4pt}        %
\begin{tabular}{|c|c|c|c|c|}
\hline
\multirow{2}{*}{Dataflow} & \multicolumn{3}{c|}{External DRAM} & Internal CIM \rule{0pt}{2.5ex} \\ \cline{2-5}
                           & Input & Weight & Output & Weight update \rule{0pt}{2.5ex} \\ \hline
IS                         & $MN$ & $\frac{M}{m}\!\times\!NK$ & $\frac{N}{n}\!\times\!MK$ & $\frac{M}{m}\!\times\!NK$ \\ \hline
WS                         & $\frac{K}{k}\!\times\!MN$ & $NK$ & $\frac{N}{n}\!\times\!MK$ & $NK$ \\ \hline
IS-OS\cite{2025_NSFormer}  & $MN$ & $\frac{M}{m}\!\times\!NK$ & $MK$ & $\frac{M}{m}\!\times\!NK$ \\ \hline
WS-OS\cite{2025_NSFormer}  & $\frac{K}{k}\!\times\!MN$ & $NK$ & $MK$ & $\frac{M}{m}\!\times\!NK$ \\ \hline
WS-OCS                     & $\frac{K}{k}\!\times\!(M-m)N$ & $NK$ & $MK$ & $NK$ \\ \hline
\end{tabular}
\end{table}

\subsection{Dataflow with Weight-stationary output-column-stationary (WS-OCS)}

To reduce external DRAM bandwidth, digital AI accelerators have adopted two common dataflows, input-stationary output-stationary (IS-OS) and weight-stationary output-stationary (WS-OS), to maximize data reuse, as shown in Fig.~\ref{5_OCS_dataflow}.
However, IS-OS and WS-OS are unsuitable for CIM architectures due to frequent weight updates.
To address this, Fig.~\ref{5_OCS_dataflow}(c) presents the proposed weight-stationary output-column-stationary (WS-OCS) dataflow, where block weights are loaded and computed column-wise with input data, while partial sums are buffered internally.
The input data is then scanned along the N dimension to compute column outputs, repeating this process for all weight blocks to produce the final output.

Table~\ref{tb01_WS_OCS_dataflow} summarizes external DRAM access and internal CIM weight-update behavior for IS, WS, IS-OS, WS-OS, and WS-OCS dataflows.
To minimize external DRAM traffic, WS-OCS employs an input-reuse buffer, reducing input DRAM access from $\frac{K}{k} \times MN$ to $\frac{K}{k} \times (M - m)N$ compared with WS-OS.
Moreover, WS-OCS markedly lowers CIM weight-update latency and volume (from $\frac{M}{m} \times NK$ to $NK$) and further reduces weight-buffer capacity by replacing weights only after all corresponding inputs are fully processed.

\begin{figure}[tb]
\centering
\includegraphics[width=1.0\linewidth]{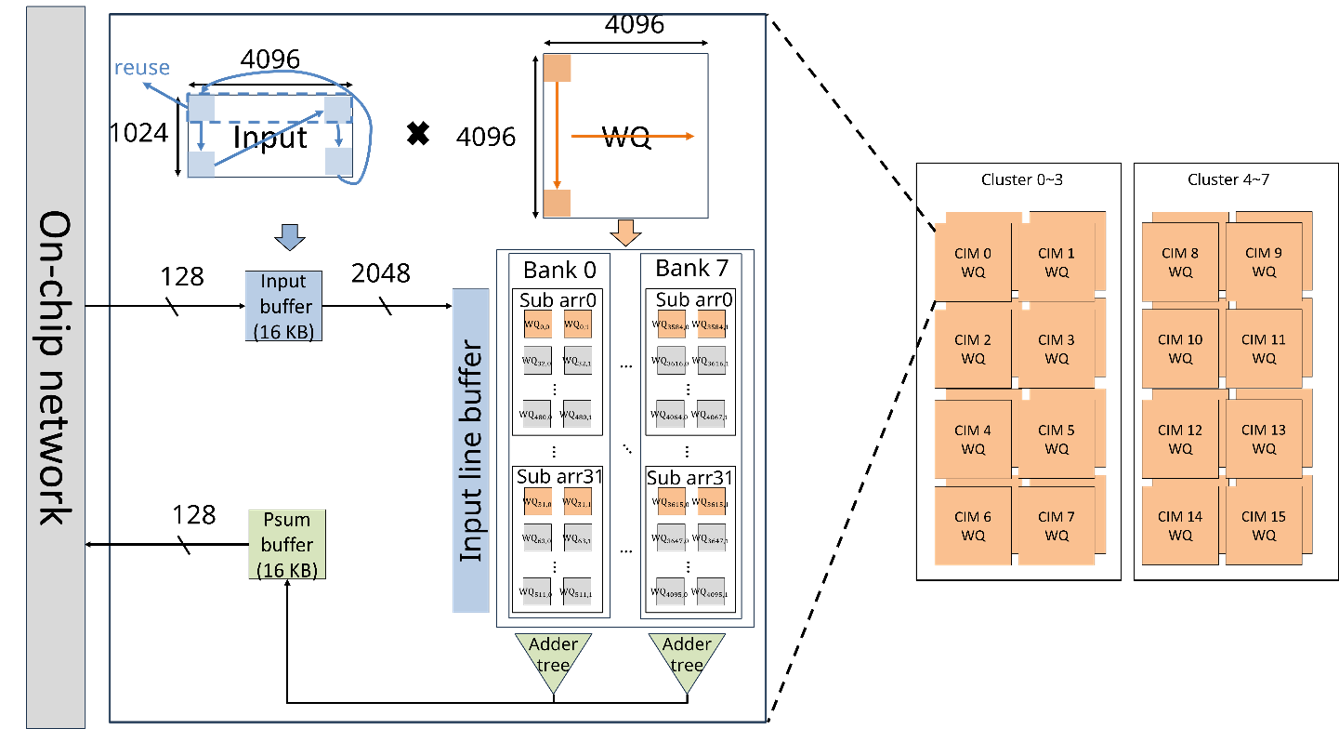}
\caption{Mapping of the WS-OCS dataflow onto the CIM macro.}
\label{6_OCS_dataflow_mapping}
\end{figure}

Fig.~\ref{6_OCS_dataflow_mapping} illustrates the CIM macro mapping for LLM under the WS-OCS dataflow. In this scheme, Q weight blocks are distributed across the eight CIM clusters, while the input line buffer is shifted through the input-reuse buffer. The partial-sum buffer stores the intermediate column outputs, and the complete output results are obtained by scanning the input data along the N dimension and subsequently written back to DRAM.

\subsection{Low-Latency nonlinear operator fusion in CIM Macro}

Online Softmax and group RMSNorm have been proposed to reduce global dependency latency on systolic arrays \cite{2025_systolicattention}. However, CIM architectures still suffer from high power consumption and low efficiency when handling exponential and division operations. Thus, a 64-segment look-up table (LUT)-based group Softmax is adopted to reduce nonlinear dependency latency and prevent overflow and underflow, as shown in (\ref{eq01_group_softmax}).

\begin{equation}
\text{Softmax}(x_i) \approx 
\frac{\text{LUT}_{64}(x_i - \max_{g \in G} x_g)}
{\sum_{g \in G} \text{LUT}_{64}(x_g - \max_{g \in G} x_g)}, 
\quad i \in G
\label{eq01_group_softmax}
\end{equation}

\begin{table*}[htb]
\caption{Summary of performance and comparison with other designs.}
\label{tb02_HW_comparison}
\centering
\begin{tabular}{|c|c|c|c|}
\hline                          & VLSI’25\cite{2025_LLM_CIM}                    & VLSI'25\cite{2025_CELLA}                      & \textbf{This work}                                            \\ \hline
Technology                      & 28nm                                          & 28nm                                          & \textbf{22nm}                                                 \\ \hline
Architecture                    & SRAM-based digital CIM                        & SRAM-based digital CIM                        & \textbf{SRAM-based digital CIM}                               \\ \hline
Macro size (KB)                 & 12                                            & 128                                           & \textbf{256}                                                  \\ \hline
Supply Voltage (V)              & 0.65                                          & 0.7                                           & \textbf{0.8}                                                  \\ \hline
Frequency (MHz)                 & 85-410                                        & 110-500                                       & \textbf{100}                                                  \\ \hline
Input precision                 & INT4 / INT8                                   & BF16                                          & \textbf{INT8 / BF16}                                          \\ \hline
Weight precision                & INT4 / INT8                                   & INT4 / INT8                                   & \textbf{INT4 / INT8 / BF16}                                          \\ \hline
Throughput (TOPS)               & \makecell{0.63 \\ (0.63\footnotemark[1])}     & \makecell{2.98 \\ (5.96\footnotemark[1])}     & \makecell{\textbf{3.28} \\ \textbf{(3.28\footnotemark[1])}}   \\ \hline
Energy efficiency (TOPS/W)      & \makecell{36.3 \\ (30.5\footnotemark[2])}     & \makecell{15.6 \\ (30.4\footnotemark[2])}     & \makecell{\textbf{42.3} \\ \textbf{(42.3\footnotemark[2])}}   \\ \hline
Model                           & LLaMa2-7B                                     & LLaMa2-7B                                     & \textbf{LLaMa2-7B}                                            \\ \hline
Prefill latency (ms)            & 19.4                                          & 7.2                                           & \textbf{4.2}                                                  \\ \hline
Decode token/s                  & -                                             & 8.51                                          & \textbf{26.87}                                                \\ \hline
Support nonlinear operation     & \makecell{\checkmark  (not support fusion)}   & -                                             & \makecell{\textbf{\checkmark}  \textbf{(support fusion)}}     \\ \hline
Weight update optimization      & -                                             & -                                             & \textbf{\checkmark}                                           \\ \hline
\end{tabular}
\footnotemark[1]{%
\begin{minipage}[t]{0.95\textwidth}
Normalized operations (in equivalent 8-bit ops) = operations$\times$ $\frac{\text{input precision}}{\text{8 bit}}$ $\times$ $\frac{\text{weight precision}}{\text{8 bit}}$
\end{minipage}%
}
\footnotemark[2]{%
\begin{minipage}[t]{0.95\textwidth}
Normalized energy efficiency (in equivalent 8-bit precision) = energy efficiency$\times$ $\frac{\text{input precision}}{\text{8 bit}}$ $\times$ $\frac{\text{weight precision}}{\text{8 bit}}$ $\times \left(\frac{\text{process}}{\text{22 nm}}\right) \times \left(\frac{\text{voltage}}{\text{0.8 V}}\right)^2$
\end{minipage}%
}
\end{table*}

\begin{figure}[tb]
\centering
\includegraphics[width=0.8\linewidth]{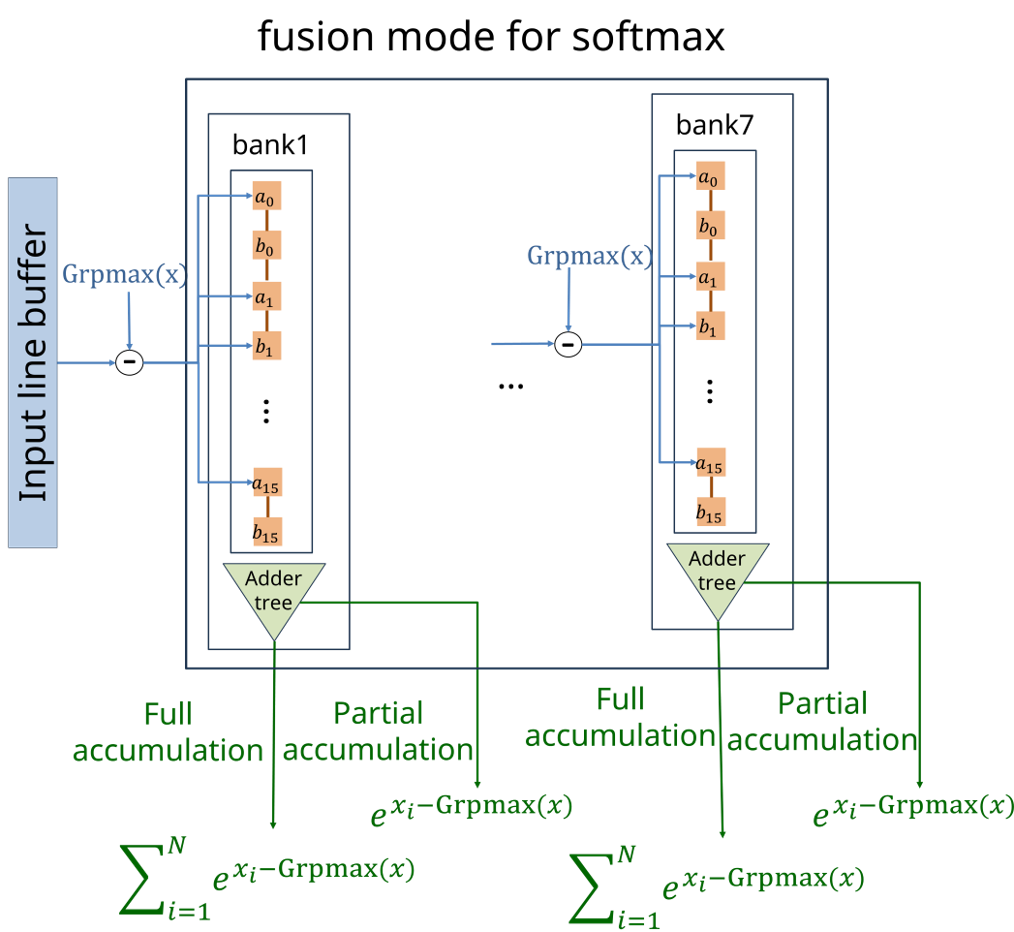}
\caption{Nonlinear group softmax fusion in the CIM macro supports full and partial accumulation to compute exponentials in parallel, improving overall efficiency.}
\label{8_nonlinear_fusion}
\end{figure}

Prior CIM implementations~\cite{2025_LLM_CIM} executed softmax inefficiently because they only supported full accumulation, which significantly reduced CIM utilization. To address this issue, we propose nonlinear operator fusion in the CIM macro, which not only supports full accumulation for computing the sum of exponentials, but also enables partial accumulation to perform exponentiation in parallel, thereby improving overall CIM efficiency.

Fig.~\ref{8_nonlinear_fusion} illustrates the group softmax fusion in the CIM macro. To minimize global dependencies, the input is first offset by the group maximum.
The approximate coefficients a and b are then stored in the CIM macro using a 64-segment LUT.
Finally, the adder tree performs both full and partial accumulation to efficiently compute the exponent sum and exponentiation.
Additionally, high-precision FP16 is used to compute softmax, reducing accuracy loss.

RMSNorm\cite{2019_RMSNorm} applies group RMS to reduce global latency.
To further save computation time and mitigate accuracy loss, the synchronization to the global RMS is performed together with the $\gamma$ scaling, as shown in (\ref{eq02_group_RMSNrom}).
\begin{equation}
\text{RMSNorm}(x_i) =
\frac{x_i}{\sqrt{\frac{1}{|G_m|} \sum_{j \in G_m} x_j^2 + \epsilon}} 
\cdot \gamma_i, \quad i \in G_m
\label{eq02_group_RMSNrom}
\end{equation}

\section{Experimental result} 
RCW-CIM is implemented using TSMC 22 nm CMOS at 100 MHz, achieving 3.28 TOPS and 42.3 TOPS/W.
The Llama2-7B model employs INT4 weights, INT8 activations, and FP16 for nonlinear functions.
Table~\ref{tb02_HW_comparison} summarizes performance comparisons.
While \cite{2025_LLM_CIM} supports nonlinear functions, it lacks nonlinear operator fusion and relies on INT8, causing higher precision loss.
RCW-CIM applies nonlinear operator fusion to reduce nonlinear function latency and uses FP16 to maintain accuracy.
Moreover, \cite{2025_LLM_CIM} and \cite{2025_CELLA} do not consider weight update latency; RCW-CIM addresses this via RCW, minimizing latency.
By combining WS-OCS dataflow with RCW, RCW-CIM achieves 4.2 ms prefill latency and 26.87 decoded tokens per second for the INT4-weight Llama2 model using dual DDR5-6400 memory.

\begin{figure}[bth]
\centering
\includegraphics[width=0.9\linewidth]{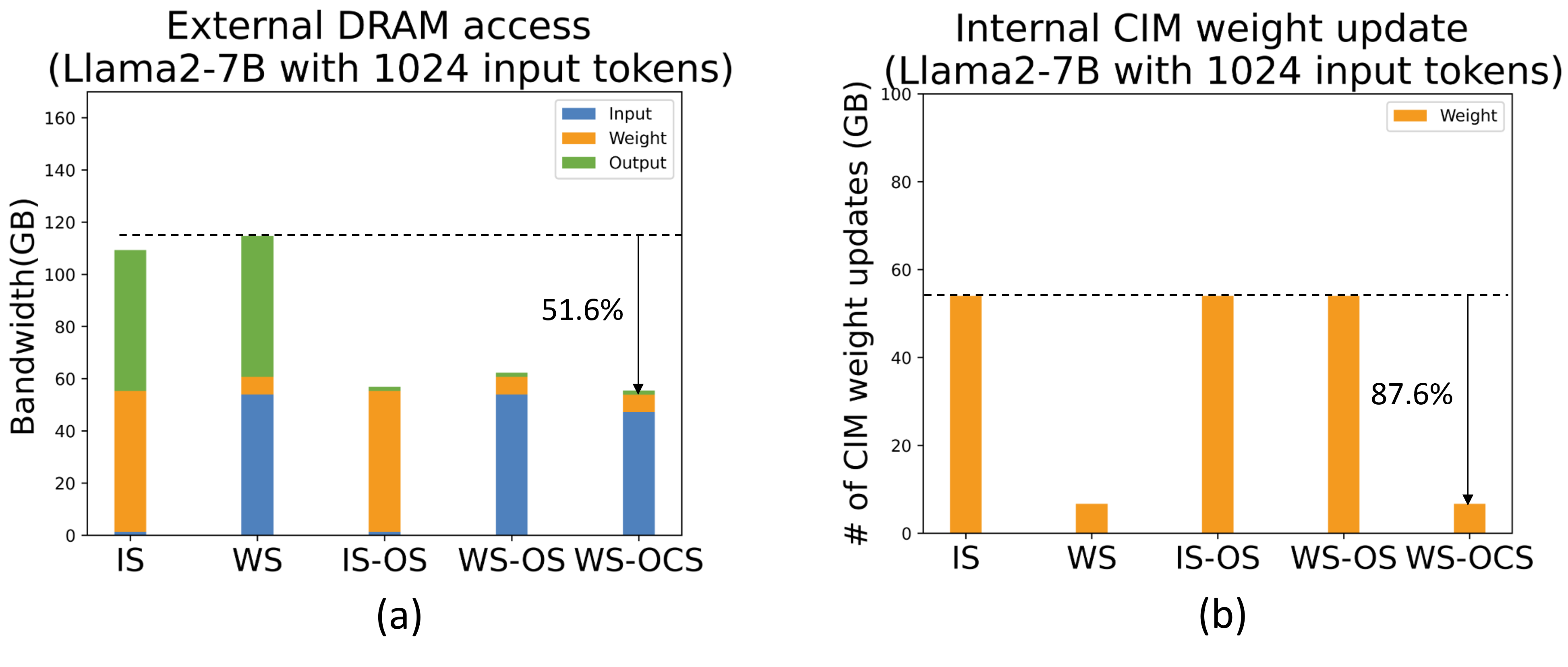}
\caption{(a) Compared with the WS dataflow, the WS-OCS scheme reduces external memory access by 51.6\% during the prefill stage with 1024 input tokens.
(b) The WS-OCS dataflow also decreases internal CIM weight-update operations by 87.6\% compared with the IS-OS and WS-OS schemes.}
\label{7_OCS_prefill_EMA}
\end{figure}

Fig.~\ref{7_OCS_prefill_EMA} analyzes the impact of the proposed approaches.
As shown in (a), WS-OCS reduces external DRAM access by 51.6\% compared with WS via partial sums and input-reuse buffers.
As shown in (b), WS-OCS further cuts internal CIM weight-update operations by 87.6\% relative to IS-OS and WS-OS through column-wise weight mapping and N-dimensional input scanning.
\begin{figure}[bt]
\centering
\includegraphics[width=0.9\linewidth]{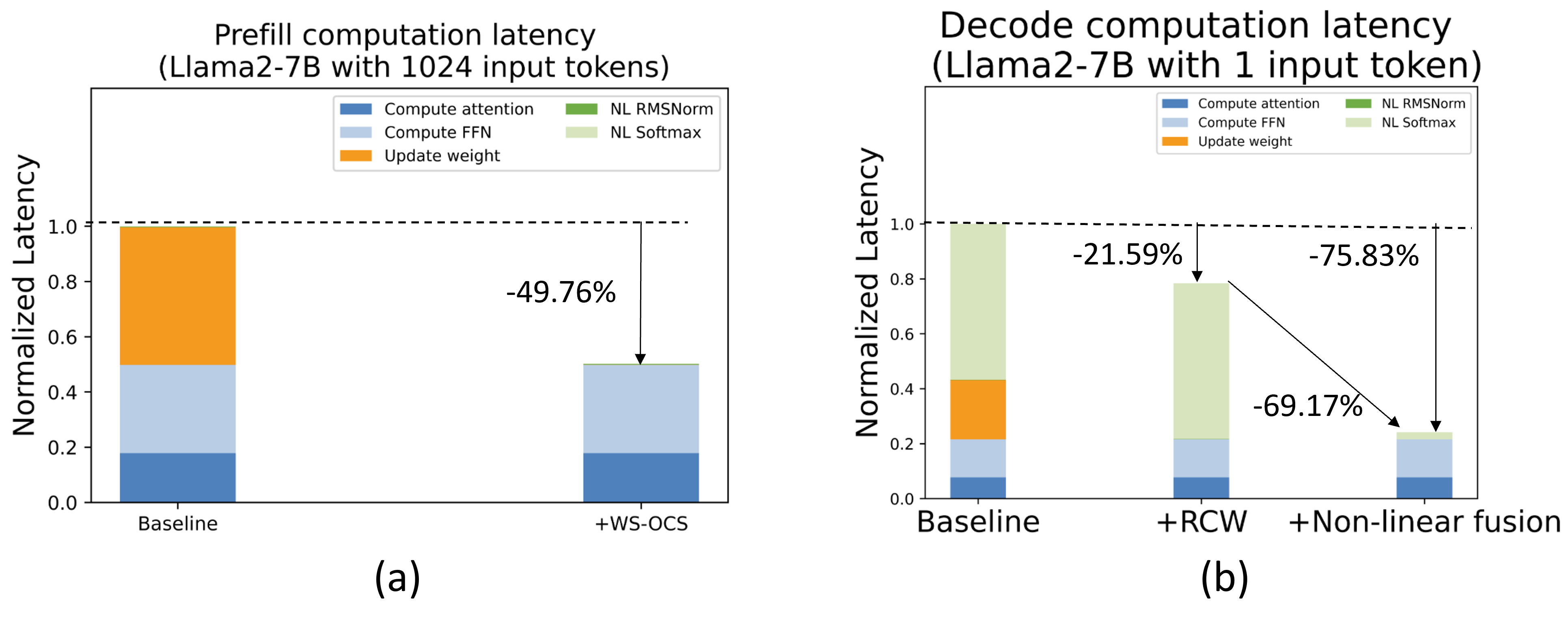}
\caption{The baseline employs multi-macro parallel weight updates to reduce weight-update latency.
Through the proposed methods, performance is further improved:
(a) The WS-OCS dataflow reduces computation latency by 49.76\% during the prefill stage.
(b) The RCW and nonlinear operator fusion modules further reduce latency by 75.83\% during the decode stage.}
\label{9_compute_latency}
\end{figure}

Fig.~\ref{9_compute_latency} shows that the proposed techniques progressively reduce computation latency: WS-OCS cuts prefill latency by 49.76\%, RCW reduces decode latency by 21.59\%, and nonlinear operator fusion combined with group Softmax and RMSNorm further lowers decode latency by an additional 69.17\%.

\section{Conclusion} 
This paper presents RCW-CIM, achieving 3.28 TOPS and 42.3 TOPS/W using TSMC 22 nm CMOS technology at 100 MHz.
RCW-CIM introduces the WS-OCS dataflow, which reduces external DRAM access by 51.6 \% and internal CIM weight updates by 87.6 \% during the prefill of 1024 tokens, leading to a 49.76 \% reduction in prefill latency.
Furthermore, the proposed RCW mitigates weight update overhead, reducing decode computing latency by 21.59 \%.
The nonlinear operator fusion technique further decreases decode latency by 69.17 \% through partial accumulation, while group Softmax and RMSNorm synchronization minimize global latency.
As a result, RCW-CIM achieves a prefill latency of 4.2 ms and a decoding throughput of 26.87 tokens/s for the INT4-weight Llama2-7B model with dual DDR5-6400 memory.

\section*{Acknowledgment}
This work was supported by the National Science and Technology Council, Taiwan, under Grant 111-2622-8-A49-018-SB, 110-2221-E-A49-148-MY3, 113-2221-E-A49-078-MY3, and 113-2640-E-A49-005.

\bibliographystyle{IEEEtran}

\bibliography{bib/ieeeBSTcontrol,bib/thesis}

\end{document}